\documentclass[manuscript]{aastex}
\usepackage{emulateapj5}
\usepackage{graphicx}
\usepackage{times}

\shortauthors{Hopkins, Rao \& Turnshek}
\shorttitle{}

\begin{document}

\title{The star formation history of damped Lyman alpha absorbers}

\author{A. M. Hopkins\altaffilmark{1}, S. M. Rao, D. A. Turnshek
}

\affil{
Dept.\ of Physics and Astronomy, University
 of Pittsburgh, 3941 O'Hara Street, Pittsburgh, PA 15260
}
\altaffiltext{1}{email ahopkins@phyast.pitt.edu}

\begin{abstract}
The local power law relationship between the surface densities of neutral
hydrogen gas and star formation rate (SFR) can be used to explore the
SFR properties of damped Lyman $\alpha$ (DLA) systems at higher redshift.
We find that while the SFR densities for DLA systems are consistent with
luminous star forming galaxies at redshifts below $z \approx 0.6$, at higher
redshifts their SFR density is too low for them to provide a significant
contribution to the cosmic star formation history (SFH). This suggests that the
majority of DLAs may be a distinct population from the Lyman break galaxies
(LBGs) or submillimeter star-forming galaxies that together dominate the SFR
density at high redshift. It is also possible that the DLAs do not trace the
bulk of the neutral gas at high redshift. The metallicity properties of DLAs
are consistent with this interpretation. The DLAs show a metal mass density
lower by two orders of magnitude at all redshifts than that inferred from the
SFH of the universe. These results are consistent with DLAs being dominated
by low mass systems having low SFRs or a late onset of star formation,
similar to the star formation histories of dwarf galaxies in the
local universe.
\end{abstract}

\keywords{galaxies: evolution --- galaxies: formation ---
 galaxies: starburst --- quasars: absorption lines}

\section{Introduction}
\label{int}

The nature of damped Ly$\alpha$ (DLA) galaxies remains an unresolved
question, although recent work has advanced our understanding
considerably. \citet{York:86} proposed that at least some DLAs were dwarf
and low surface-brightness galaxies. This has
been confirmed in a number of cases for DLAs at $z\lesssim 1$
\citep[e.g., ][]{RSL:04,Rao:03,Tur:01,RT:98}. There are alternative
suggestions that DLAs are similar to the population of Lyman break galaxies
(LBGs) at high redshift based on their star formation (SF) properties
\citep{Wol:03a,Wol:03b,Wea:05}. Here we explore the nature of the DLA
population further through their star formation history (SFH). Using the local
relation between gas and star formation rate (SFR) surface densities from
\citet{Ken:98} we compare the space densities of DLA SFRs with the cosmic
history of SFR density for luminous galaxies.

\begin{figure*}
\centerline{\rotatebox{-90}{\includegraphics[width=10cm]{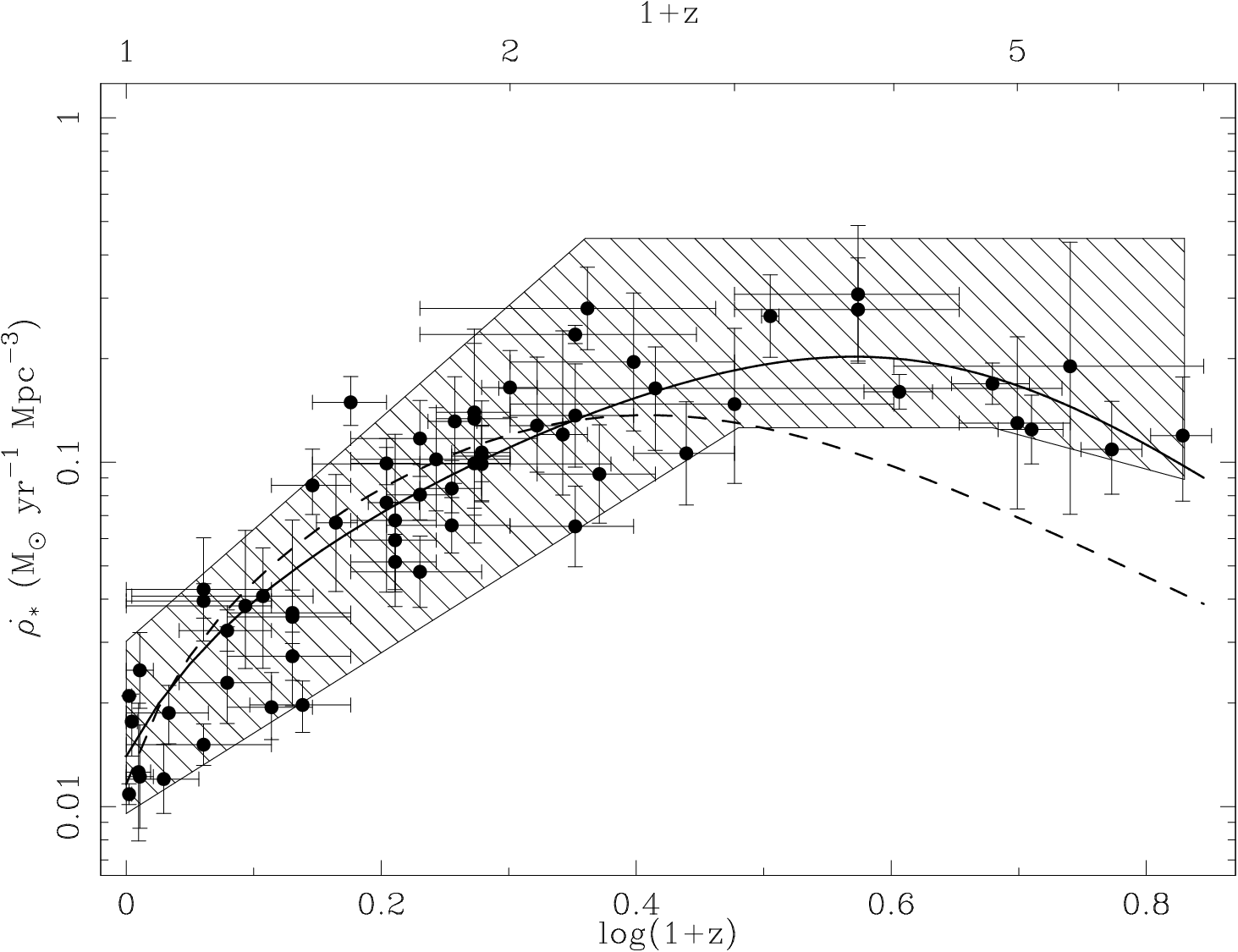}}}
\caption{The compilation of SFR density data from \citet{Hop:04}.
The dashed line is the parameterised form from
\citet{Col:01}, which is consistent with the data only for $z\lesssim2$.
At higher redshifts this curve underestimates the SFR density compared
with the measurements. The solid line is an updated fit to the data
using the analytic form from \citet{Col:01} as detailed in the text.
The hatched region, established by visual inspection,
encompasses the majority of the most reliable SFR density estimates
at all redshifts.
 \label{fig1}}
\end{figure*}

The evolution of the volume density of SFR in galaxies
has been recently summarised by \citet[his Figure~1, for example]{Hop:04},
and we reproduce those data points here in Figure~\ref{fig1}. The hatched
envelope in this figure encompasses the majority of these data, including
estimates corresponding to the ``SFR-dependent" obscuration correction,
which increases slightly the height of the upper envelope above $z\approx 1$.
This envelope can be used as a realistic bound from observational
measurements to constrain the cosmic star formation history (SFH) and its
related integral quantities, and its vertices are given in Table~\ref{tab1}.
The parameterisation of the SFH (corrected for dust obscuration) from
\citet{Col:01} has come to be relied upon by the community due to its
useful analytic form, and it is illustrated in Figure~\ref{fig1}, shown
as the dashed curve. This curve, based on the much smaller data compilation
available at that time, is consistent with the current compilation at
$z\lesssim 2$ although it significantly underestimates the SFH for higher
redshifts. Also shown as the solid line in Figure~\ref{fig1} is an updated fit
using the \citet{Col:01} analytic form, $\dot{\rho}_*=(a+bz)h/(1+(z/c)^d)$,
with $h=0.7$, and parameters $a=0.02$, $b=0.14$, $c=3.6$ and $d=3.4$.

A \citet{Sal:55} initial mass function (IMF) is assumed throughout, with
lower and upper mass limits of $0.1$ and $100\,$M$_{\odot}$.
We adopt a cosmology with $h=0.7, \Omega_M=0.3, \Omega_\Lambda=0.7$,
where $H_0=100\,h\,$km\,s$^{-1}$\,Mpc$^{-1}$.

\section{SFR density of DLAs}
\label{dlasfr}
\citet{Ken:98}, using H$\alpha$, H{\sc i}, CO and FIR measurements for
a sample of local spiral and starburst galaxies, quantifies a global
Schmidt law relating the disk-averaged surface densities of gas and SFR.
This relationship, Equation~4 of \citet{Ken:98}, can be expressed as:
\begin{equation}
\label{schmidt}
\left(\frac{\Sigma_{\rm SFR}}{\rm M_{\odot} yr^{-1} Mpc^{-2}}\right)=
(4.0\pm1.1) \times 10^{-15}
\left(\frac{\Sigma_{\rm gas}}{\rm M_{\odot} Mpc^{-2}}\right)^{1.4\pm0.15}.
\end{equation}
Since the gas surface density in DLA systems is just the column density,
this relationship, assuming it is valid at the redshifts of the DLA
absorbers \citep[as suggested by][]{Lan:02}, can be applied to estimate
the SFR in the DLA systems. First consider the H{\sc i} mass density
$\rho_{\rm HI}=(H_0/c) m_{\rm H} \int N f(N) dN$, where $N$ is the H{\sc i}
column density and $f(N)$ is the H{\sc i} column density
distribution. This quantity is most commonly used in the
calculation of $\Omega_{\rm DLA}$:
\begin{equation}
\Omega_{\rm DLA} = \frac{H_0}{c}\frac{\mu m_{\rm H}}{\rho_{\rm crit}}
                     \int_{N_{\rm min}}^{N_{\rm max}} N f(N) dN.
\end{equation}
Here $m_{\rm H}$ is the mass of a hydrogen atom, $\rho_{\rm crit}$ is the
critical density, and $\mu=1.3$ is the factor commonly used
to incorporate helium as well as hydrogen in the estimate of the DLA neutral
gas density \citep[e.g.,][]{RT:00}. To calculate the SFR density, then,
it is necessary to integrate the SFR surface density (derived from the column
density using Equation~\ref{schmidt}) multiplied by the number of systems
$f(N)$ of the corresponding column density, over $N$:
\begin{equation}
\label{rhostarfn}
\dot{\rho}_* = \frac{H_0}{c} 4.0\times10^{-15} m_{\rm H}^{1.4}
               \int_{N_{\rm min}}^{N_{\rm max}} N^{1.4} f(N) dN.
\end{equation}
This gives $\dot{\rho}_*$ in units of ${\rm M_{\odot} yr^{-1} Mpc^{-3}}$
for $m_{\rm H}$ in units of ${\rm M_{\odot}}$, $N$ in units of Mpc$^{-2}$ and
$f(N)$ in units of Mpc$^2$. For our assumed cosmology, this can be rewritten as
\begin{equation}
\dot{\rho}_* = 4.29\times10^{-30}
               \int_{N_{\rm min}}^{N_{\rm max}} N^{1.4} f(N) dN
\end{equation}
with $N$ and $f(N)$ in traditional units of cm$^{-2}$ and cm$^2$,
respectively. The column density distribution function is often
parameterised as $f(N)=B N^\beta$, giving
\begin{equation}
\dot{\rho}_* = 4.29\times10^{-30} \frac{B}{2.4+\beta}
            \left(N_{\rm max}^{2.4+\beta} - N_{\rm min}^{2.4+\beta}\right).
\end{equation}

The quantity $f(N)$ requires detection of many absorber systems to
be reliably measured, and is thus difficult to measure as a function of
redshift. Conveniently, though,
\begin{equation}
\label{sighi}
\int N f(N) dN = \frac{c}{H_0} \frac{n_{\rm DLA}}{dX/dz}
                 \langle N_{\rm HI}\rangle
\end{equation}
where $n_{\rm DLA}$ is the redshift distribution of the DLA systems $dn/dz$,
and the ``absorption distance" $dX/dz=(c/H_0)(1+z)^2/E(z)$, with
$E(z)=(\Omega_M(1+z)^3 + \Omega_k(1+z)^2 + \Omega_{\Lambda})^{0.5}$. This
allows the mean column density $\langle N_{\rm HI}\rangle$ to be explicitly
used, rather than the integral of $f(N)$, when estimating mass
densities of DLA gas:
\begin{equation}
\rho_{\rm HI} = \frac{n_{\rm DLA}}{dX/dz} m_{\rm H} \langle N_{\rm HI}\rangle.
\end{equation}
We could analogously write
\begin{equation}
\label{rhostaravsig}
\dot{\rho}_* = \frac{n_{\rm DLA}}{dX/dz} \langle \Sigma_{\rm SFR} \rangle
\end{equation}
using Equation~\ref{rhostarfn} to infer that
\begin{equation}
\label{sigsfr}
\langle \Sigma_{\rm SFR} \rangle = \frac{H_0}{c} \frac{dX/dz}{n_{\rm DLA}}
 4.0\times10^{-15} m_{\rm H}^{1.4} \int_{N_{\rm min}}^{N_{\rm max}}
 N^{1.4} f(N) dN.
\end{equation}
Alternatively, from Equation~\ref{rhostaravsig} we could
invoke Equation~\ref{schmidt} to infer
\begin{eqnarray}
\dot{\rho}_* & = & \frac{n_{\rm DLA}}{dX/dz} 4.0\times10^{-15} \langle \Sigma_{\rm gas} \rangle^{1.4} \nonumber \\
\label{rhostarn}
             & = & \frac{n_{\rm DLA}}{dX/dz} 4.0\times10^{-15} (m_{\rm H}\langle N_{\rm HI}\rangle)^{1.4}
\end{eqnarray}
(specifying that $m_{\rm H}\langle N_{\rm HI}\rangle$ is in units of
${\rm M_{\odot} Mpc^{-2}}$). Here we are making the coarse assumption
that the mean column density can be used to infer the mean SFR surface
density for the whole population of DLAs. We will show that this assumption
is not strictly true, but it does provide a useful approximation, and allows
estimates of $\dot{\rho}_*$ to be made in the absence of robustly measured
$f(N)$ distributions.

In order to evaluate the reliability of Equation~\ref{rhostarn}, we need
to show that the ratio
\begin{equation}
R = \frac{\langle \Sigma_{\rm SFR} \rangle}
 {4.0\times10^{-15} (m_{\rm H}\langle N_{\rm HI}\rangle)^{1.4}} \approx 1.
\end{equation}
From Equations~\ref{sighi} and \ref{sigsfr} we have
\begin{equation}
R = \left(\frac{c}{H_0} \frac{n_{\rm DLA}}{dX/dz}\right)^{0.4}
  \frac{\int N^{1.4} f(N) dN}{\left[\int N f(N) dN\right]^{1.4}}
\end{equation}
and assuming $f(N)=B N^{\beta}$,
\begin{equation}
R = \left(\frac{c}{H_0} \frac{n_{\rm DLA}}{dX/dz}\right)^{0.4}
   \frac{(2+\beta)^{1.4}}{B^{0.4}(2.4+\beta)}
   \frac{(N_{\rm max}^{2.4+\beta} - N_{\rm min}^{2.4+\beta})}
   {\left(N_{\rm max}^{2+\beta} - N_{\rm min}^{2+\beta}\right)^{1.4}}.
\end{equation}
Now, with values of $B$, $\beta$ for $f(N)$ distributions at given
redshifts, along with $n_{\rm DLA}$, we can evaluate $R$ corresponding to
integrals over appropriate ranges of $N$. At intermediate redshift,
$0.1 \lesssim z \lesssim 1.6$, $\beta=-1.40$, $B=10^{6.40}$, and
$n_{\rm DLA}\approx0.1$ \citep{Rao:05b}, integrating over the range
$20.3 \le \log(N) \le 21.8$ gives $R=1.4$. At high redshift,
$2\lesssim z \lesssim 4$, $\beta=-1.79$, $B=10^{14.8}$, and
$n_{\rm DLA}\approx0.25$, and integrating over the same range gives $R=1.3$.
At $z=0$, $f(N)$ has been determined by \citet{Rya:03} and parameterised
as a broken power-law, with $\beta=-1.4$ for $19.6 \le \log(N) \le 20.9$
and $\beta=-2.1$ for $20.9 \le \log(N) \le 21.6$. The corresponding
normalisations are $B=10^{6.08}$ and $B=10^{21.46}$. Locally,
$n_{\rm DLA}=0.046$ \citep{Rya:05}. Then integrating over the full range
$19.6 \le \log(N) \le 21.6$ gives $R=1.6$.

The advantage of Equation~\ref{rhostarn} is that it allows published values
of $\Omega_{\rm DLA}$ and $n_{\rm DLA}$ to be used to derive $\dot{\rho}_*$.
This also allows a higher redshift resolution to be obtained for
$\dot{\rho}_*$, as $\Omega_{\rm DLA}$ and $n_{\rm DLA}$ have been estimated
for more redshift bins than the number for which $f(N)$ has been accurately
parameterised. For an arbitrary cosmology:
\begin{equation}
\label{rhostaromega}
\dot{\rho}_*^{c2} = 4.0\times10^{-15} \frac{n_{\rm DLA}}{dX^{c2}/dz}
\left(\frac{dX^{c1}/dz}{n_{\rm DLA}} \frac{\Omega_{\rm DLA}^{c1}}{\mu}
\rho_{\rm crit}^{c1} \right)^{1.4}
\end{equation}
where $c1$ and $c2$ represent the measurements using the initial and desired
cosmological parameters, respectively.
The term in the brackets is just $m_{\rm H}\langle N_{\rm HI}\rangle$ derived
from the published parameters. Where no cosmological conversion is
necessary, this simplifies to
\begin{equation}
\label{rhostaromega2}
\dot{\rho}_* = 4.0\times10^{-15} \left(\frac{dX/dz}{n_{\rm DLA}}\right)^{0.4}
  \left(\frac{\Omega_{\rm DLA}}{\mu} \rho_{\rm crit} \right)^{1.4}.
\end{equation}

We find that Equation~\ref{rhostaromega2} actually gives different estimates
for $\dot{\rho}_*$ than the equivalent calculation from explicitly using
Equation~\ref{sighi} prior to applying Equation~\ref{rhostarn}.
This is in the sense that Equation~\ref{rhostaromega2} produces larger
values of $\dot{\rho}_*$, and we explain this by noting that the observed
$f(N)$ measurements (used to estimate $\Omega_{\rm DLA}$) deviate somewhat
from a simple power law. The $\dot{\rho}_*$ estimates from
Equation~\ref{rhostaromega2} are coincidentally very similar to the estimates
inferred from explicitly evaluating Equation~\ref{rhostarfn}. What this means
is that if the true $f(N)$ distribution really is a power law, and this is
well-modelled by the assumed parameters detailed above, then the results
inferred from Equation~\ref{rhostaromega2} are appropriate estimates of
the true $\dot{\rho}_*$ values. If the $f(N)$ distributions instead do
have significant deviations from a power law, then the values for
$\dot{\rho}_*$ from Equation~\ref{rhostaromega2} will need to be scaled
by the appropriate $R$ value above to reliably approximate $\dot{\rho}_*$.

\begin{figure*}
\centerline{\rotatebox{-90}{\includegraphics[width=10cm]{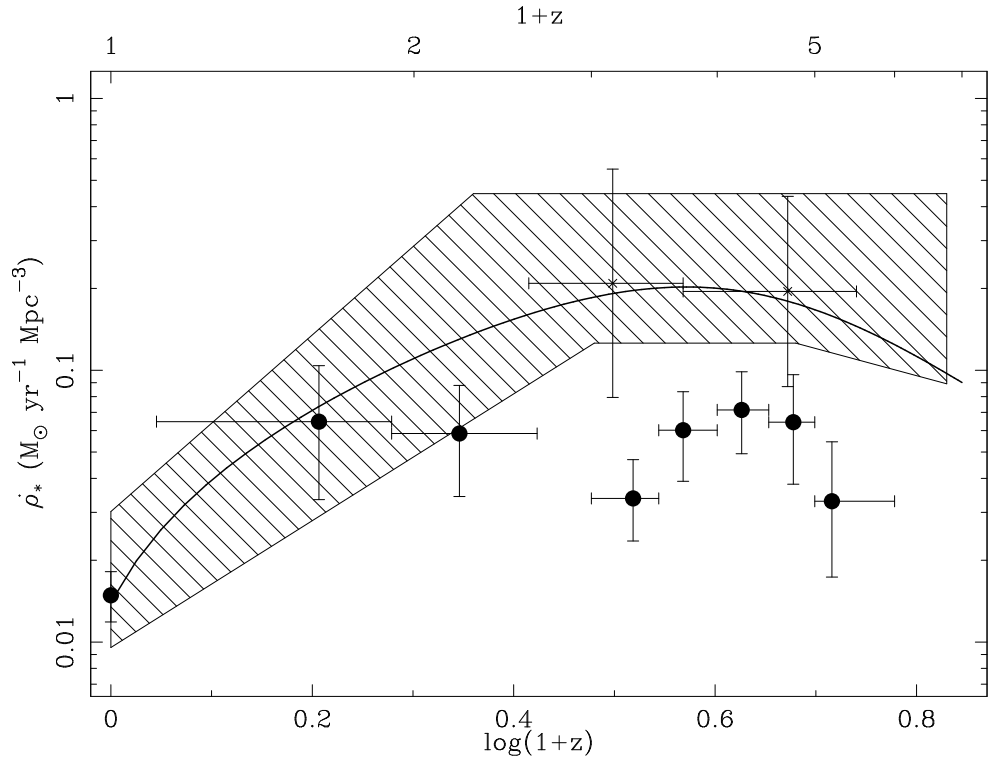}}}
\caption{The SFR density of DLA systems (filled circles) as a function
of redshift, derived as described in the text. The crosses are the
estimates from \citet{Wol:03b} using the C\,{\sc ii}$^*$ method.
Solid line and hatched region as in Figure~\protect\ref{fig1}.
 \label{fig2}}
\end{figure*}

Bearing this in mind, we obtain estimates for $\dot{\rho}_*$ at moderate to
high redshift using Equation~\ref{rhostaromega2}, with values of
$\Omega_{\rm DLA}$ and $n_{\rm DLA}$ taken from \citet{Rao:05b} and
\citet{Pro:04}. At $z=0$, where $f(N)$ is more reliably determined,
we use Equation~\ref{rhostarfn} with $f(N)$ as parameterised by \citet{Rya:03}.
These are shown in Figure~\ref{fig2} and compared with the envelope defined
in \S\ref{int}, corresponding to $\dot{\rho}_*$ estimates from luminous
galaxies.
This result is explored in detail in \S\ref{disc} below.

\section{Stellar mass density}
\label{rhostar}
\begin{figure*}
\centerline{\rotatebox{-90}{\includegraphics[width=10cm]{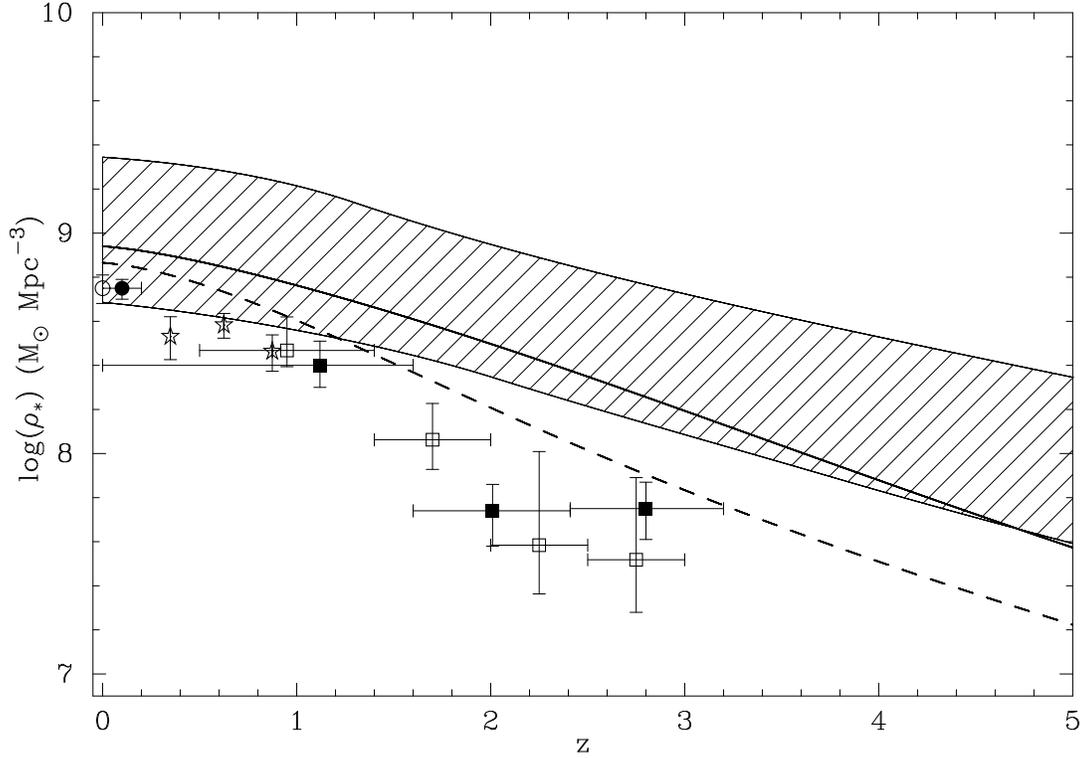}}}
\caption{The hatched region shows the mass density in stars derived from
integrating the SFH corresponding to the hatched region in
Figure~\ref{fig1}. The dashed and solid lines come from integrating the
SFH of \citet{Col:01}, and the new fit using that form, respectively.
The open circle is the local stellar density from \citet{Col:01}; the filled
circle and filled squares represent the SDSS and FIRES data, respectively,
from \citet{Rud:03}, scaled such that the SDSS measurement is consistent with
that from \citet{Col:01}; the open stars are from \citet{Bri:00}; and the
open squares are from \citet{Dic:03}.
 \label{fig3}}
\end{figure*}

The time integral of $\dot{\rho}_*$ establishes the stellar mass density,
$\rho_*(z)$, as a function of redshift. The envelope showing this in
Figure~\ref{fig3} is derived from the hatched region in Figure~\ref{fig1},
and a mass lock-up fraction of 0.72 has been used \citep{Col:01},
appropriate for the \citet{Sal:55} IMF. In this diagram, we assume
the redshift of formation is $z_f=10$, although the results are not strongly
sensitive to this choice.
The \citet{Col:01} parameterisation, along with our new fit, are similarly
integrated. The differences between the \citet{Col:01} curve and the high-z
measurements in Figure~\ref{fig1} appear here as distinct predictions
of the stellar mass density for $z\gtrsim 1$.

Several independent measurements of $\rho_*(z)$ are shown for comparison,
after conversion to our adopted cosmology where necessary. At low redshift
($z\lesssim1$), the measurements are encouragingly consistent with
the integral of the SFH \citep{Bri:00,Col:01,Rud:03}, albeit toward the
lower limit of the envelope. The data from \citet{Bri:00} have
been increased by a factor 1.25, as in earlier comparisons
\citep{Dic:03,Rud:03}, to account for their claimed $\approx 80\%$
completeness. Above $z\approx 1.5$, there is a clear discrepancy between
the SFH integral and the measurements from
\citet{Rud:03} and \citet{Dic:03}.
It is important to emphasise that the \citet{Sal:55} IMF has been used
in all measurements shown here, since the resulting $\rho_*$
values are strongly sensitive to the assumed IMF.

The data of \citet{Dic:03} are shown with uncertainties representing the
larger of their $1\,\sigma$ or systematic uncertainties for each point. Their
mass estimates are based on SED template fitting to a Hubble Deep Field (HDF)
photometric sample, with upper limits derived by invoking a two-component
model that includes a maximally old, but unobscured stellar
population. They comment that arbitrarily large masses could in principle be
derived by freely adding extinction to this component. Although this aspect
is not further pursued by those authors, this possibility may not be
unreasonable. Indeed \citet{Dun:03} have shown, from a dust-mass
function estimate based on an analysis of submillimeter sources,
that there is significant mass density in dust at $z\approx 2.5$, at least
comparable to the mass density in stellar objects at that epoch.
This suggests that heavily obscured old stellar
populations might well be significant, and it is not unreasonable that
the estimates from the HDF may be low by at least a factor of two. This
would bring the upper limits of their uncertainties to within the lower region
of the SFH integral envelope. The FIRES measurements of \citep{Rud:03}
are scaled to account for incomplete sampling, assuming that the ratio
between their SDSS derived estimate at $z=0.1$ and the measurement
of \citet{Col:01} applies to the higher redshift data. This type of
correction is highly sensitive to assumptions about the evolution of
the faint end of the luminosity function, and it is again not unreasonable
that the total mass has been underestimated.

This discrepancy between the observed values for $\rho_*(z)$ and the
integral of the SFH has been explored in some detail by \citet{Nag:04a}.
They find that simulations and theoretical models produce a stellar mass
density at $z\approx3$ higher than observations, but consistent with the SFH.
They interpret this to suggest that observations might be missing
almost half the stellar mass at high redshift, citing incomplete
galaxy population sampling, and/or cosmic variance affecting the
surveys that examine only small fields of view. Encouragingly,
the evolution of $\rho_*(z)$ found here from the integral of the observed
SFH is consistent with that from the simulations described by \citet{Nag:04a}.
It is important to verify that the envelope derived from the integral
of the SFH is consistent with independently measured data for $\rho_*(z)$,
for two reasons. First, to ensure that the SFH envelope itself is a robust
constraint on the true SFH, and second because we are about to employ
it to explore the metal mass density as a function of redshift. The
latter reason is worth emphasising particularly, as there are relatively
few independent measurements of this quantity at high redshift to
serve as consistency checks on this aspect of the analysis.

\section{Metal mass density}
\label{rhoz}
For some time it has been recognized that the low mean cosmic metallicities
observed in DLAs are inconsistent with observed stellar metallicities.
For example, initial recognition of this so-called ``missing metals'' problem
led \citet{Lan:95} to discuss it in terms of the ``cosmic G-dwarf problem''
in the context of a ``closed-box'' scenario for galaxy formation. In
particular, it was clear that if DLAs traced only luminous disk galaxies,
their mean cosmic metallicities at moderate to high redshift should be
nearly an order of magnitude higher than the observed values of $\approx0.1$
solar. Higher observed metallicities would be required to match the typical
metallicities of solar-type stars in the Milky Way disk since they were formed
5 Gyrs ago. As a solution to this problem, \citet{Lan:95} proposed that
DLAs trace not only the evolution of galactic disks but also the evolution
of galactic spheroids. In part this conclusion was motivated by the
supposed constraint that the cosmological DLA gas mass density at $z=3.5$
was converted to the mass density of stars at $z=0$; but this constraint is
no longer required by the data. More recent suggestions that also lead
to the expectation that DLA metallicities should be lower include the
possibilities of significant neutral gas in dwarf galaxies and low surface
brightness galaxies \citep[e.g.][]{Rao:03} as well as significant neutral
gas at large galactocentric distances coupled with reasonable metallicity
gradients \citep{Chen:05}. At the time of the \citet{Lan:95} study, however,
there were no measurements of the cosmic SFH to further constrain the
problem. Now, with better measurements of the stellar and gaseous components
as a function of redshift (\S\ref{dlasfr}), the nature of the missing metals
problem can be defined more clearly.

The problem has also recently been explored by \citet{Dun:03} through the
partitioning of metallicity in the various components of the Universe at
$z=0$ and $z=2.5$. They conclude that the DLAs at high redshift are not the
same population as the dusty, highly star-forming submillimeter galaxies,
which contain the majority of the metals at this epoch (and possibly earlier).

\begin{figure*}
\centerline{\rotatebox{-90}{\includegraphics[width=9cm]{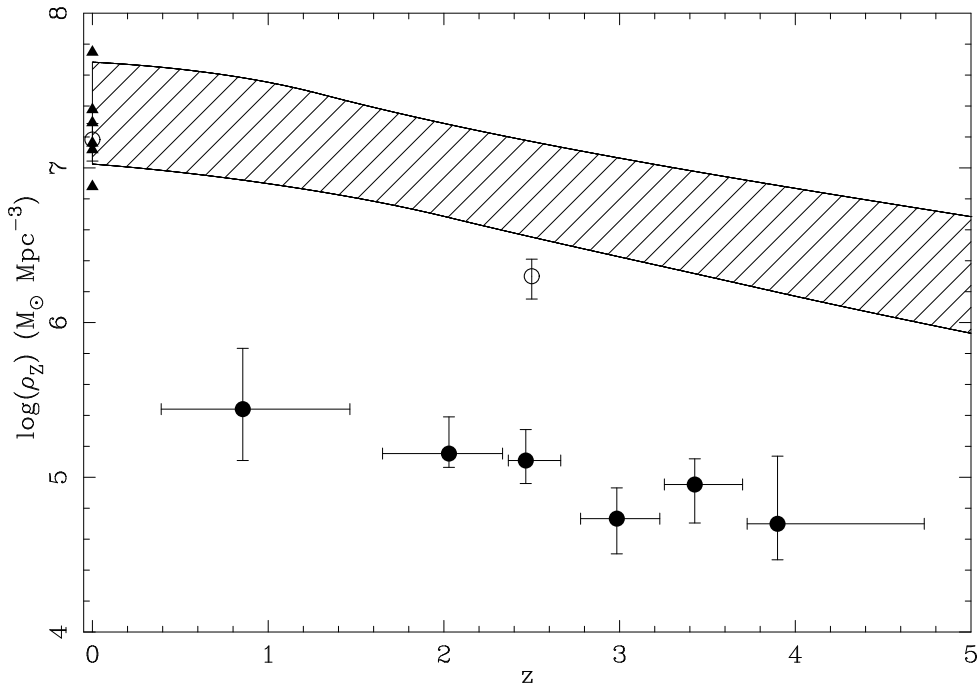}}}
\caption{The hatched region shows the mass density in metals derived from
the SFH corresponding to the hatched region in Figure~\ref{fig1}.
The filled circles are the DLA measurements, from \citet{Rao:05a};
the triangles at $z=0$ are the data from Table~7 of \citet{CM:04};
the open circles are from Table~1 of \citet{Dun:03}.
 \label{fig4}}
\end{figure*}

In the context of the current study, the metal mass density, $\rho_Z$, as a
function of redshift can be established from the SFH, since $\dot{\rho}_*$
is related to $\dot{\rho}_Z$ \citep[e.g.,][]{Mad:96}. More recent stellar
population synthesis results \citep{BC:03} indicate this relationship is
$\dot{\rho}_* = 63.7 \dot{\rho}_Z$ \citep[see, e.g.,][]{Con:03}, and the
$\rho_Z$ so derived is shown in Figure~\ref{fig4}. The local value of
$\rho_Z$ can be compared with the compilation of \citet{CM:04}, who favor a
value of $1.31\times10^7\,M_{\odot}$Mpc$^{-3}$, toward the low end of the
range. Values at $z=0$ and $z=2.5$ from \citet{Dun:03} are also shown,
indicating that the evolution in $\rho_Z$ from the SFH is reasonably
consistent with that estimated from the dusty submillimeter galaxy
population. The mass density in metals inferred from the DLAs, also
shown in Figure~\ref{fig4}, give a significantly different result in
comparison to the estimate from the SFH. To determine the mass density
in metals we use the metallicity measurements from \citet{Rao:05a}
relative to a solar metallicity mass fraction of $Z_{\odot}=0.02$
and $\log(\Omega_{\rm DLA})=-3$ for $z>0.6$ from \citet{Rao:05b},
which corresponds to $\log(\Omega_{\rm HI})=-3.11$ or
$\rho_{\rm HI}=1.05\times10^{8}\,M_{\odot}$Mpc$^{-3}$. Thus, since
\citet{Rao:05a} find that the mean cosmic metallicity of DLAs evolves from
[M/H]$=-1.6$ at $z=4$ to [M/H]$=-0.9$ at $z=1$, we infer that the mean cosmic
metallicity measured directly from DLAs at $z>0.6$ is about two orders
of magnitude less than what is inferred from the SFH. Figure~\ref{fig4}
clearly illustrates the current status of the so-called ``missing metals"
problem in a way that does not easily lend itself to solutions involving
spheroids, dwarfs, low surface brightness galaxies, or gas at large
galactocentric distance.

Thus, based on the comparison presented in Figure~\ref{fig4}, we
suggest that the missing metals problem may in fact be the result of missing
a substantial fraction of the metal-enriched gas in DLA surveys. This gas
might be in either neutral or molecular form, but even if it were in molecular
form a DLA-size column of gas would be intercepted. In particular, the method
of identifying DLA galaxies is based on gas-cross-section selection, and
one possibility is that the neutral or molecular regions containing
most of the metals have very small gas cross sections, leading to a
situation where we are missing significant amounts of both gas and
metals. In \S\ref{disc} we discuss this possibility further.

\section{Discussion}
\label{disc}

Is it valid to apply the local relationship between SFR and gas surface
densities to DLA systems at high redshift, as suggested by \citet{Lan:02}?
It seems reasonable that the physical basis underlying the Schmidt law
\citep[and references therein]{Ken:98} would not change significantly with
time, and the broad range of gas mass densities over which the relationship
has been characterized locally support this idea. But DLA systems may be
comprised of quite different gas components compared to local star forming
galaxies. There are suggestions that very little molecular gas resides in
detected DLAs \citep[e.g.,][]{Cur:04a,Cur:04b}. Such a lack of molecular gas
would suggest that if the Kennicutt relation does not hold for DLAs, it is
likely that the overall SFRs would be even lower than that inferred here.

There is, moreover, a discrepancy between the SFH determined here and
that inferred by \citet{Wol:03b} using the C\,{\sc ii}$^*$ technique.
Even if the $R=1.3$ scaling (\S\ref{dlasfr}) is applied to our $\dot{\rho}_*$
estimates at $z>2$, this is an increase of only $0.11$\,dex, not enough to
significantly reduce the discrepancy. In the currently adopted cosmology
their ``consensus" model gives
$\dot{\rho}_*=0.21^{+0.34}_{-0.13}$ at $z=2.15$, and
$\dot{\rho}_*=0.19^{+0.24}_{-0.11}$ at $z=3.70$ (Figure~\ref{fig2}). These
high values are marginally consistent with the current result, but are more
consistent with the SFH of luminous galaxies. The results are interpreted
by those authors as evidence that the DLA population is the same as the LBGs.
What could cause such a significant difference in derived $\dot{\rho}_*$?
There are a number of possible reasons, some of which may be related to the
assumptions underlying the C\,{\sc ii}$^*$ SFR estimates. These include the
requirement of a steady state condition (cooling from C\,{\sc ii}$^*$
$\lambda=1335.7$ absorption used to infer the heating rate), as well
as assumptions regarding the abundance ratios for [Fe/Si] and [Si/H],
and grain composition and depletion ratio details for different dust models
\citep{Wol:03a}. All of these independently may be reasonable approximations,
but together it is possible that the combination of individual uncertainties
introduces an overall uncertainty larger than that accounted for in the formal
error analysis. There are also possible issues arising from observational
biases and effects, if the absorption lines are marginally saturated, for
example, and from the inevitable bias toward measuring the largest
contribution to the DLA gas cross section. Conversely, the $\dot{\rho}_*$
estimates from the H{\sc i} column densities rely on the single, reasonable
assumption that the Kennicutt relation is valid for the DLA systems.
We note that the recent estimate of the external ultraviolet (UV) background
radiation \citep{ME:05} is higher than assumed by \citet{Wol:03a}, and
would lead to lower SFR densities inferred from the C\,{\sc ii}$^*$ method.
In a more recent analysis \citep{Wol:05}, using a UV background estimate
consistent with that of \citet{ME:05}, the C\,{\sc ii}$^*$ estimates of
$\dot{\rho}_*$ are indeed found to be lower than those of \citet{Wol:03b},
and now appear highly consistent with the results derived here.

\begin{figure*}
\centerline{\rotatebox{-90}{\includegraphics[width=9cm]{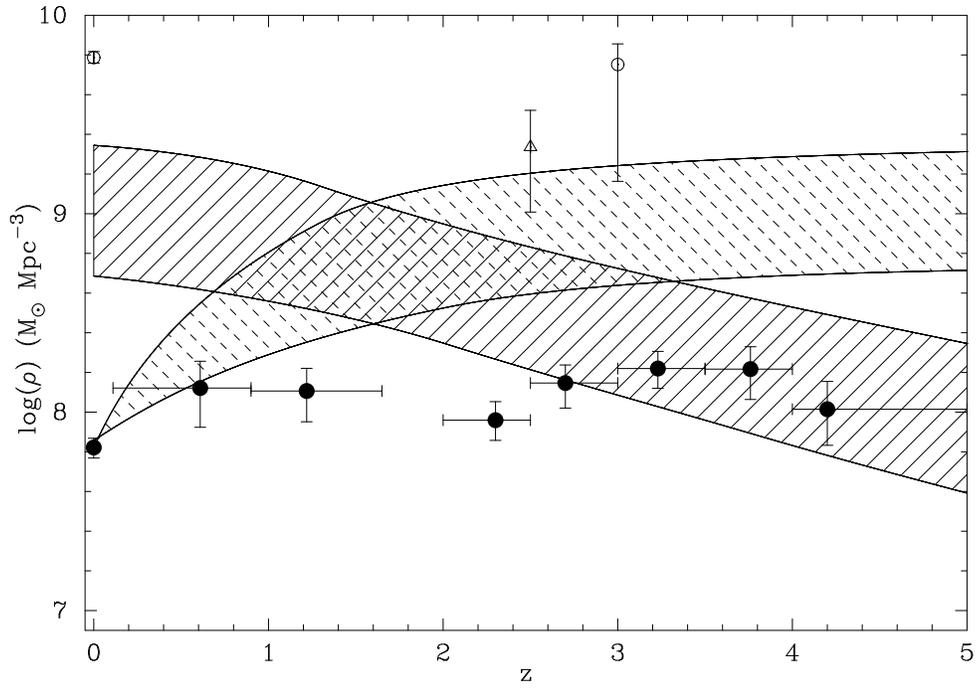}}}
\caption{The hatched region shows the evolution of the stellar mass density,
as in Figure~\ref{fig3}. The region hatched with dashed lines is
$\rho_{\rm gas}(z) = \rho_*(0) + \rho_{\rm gas}(0) - \rho_*(z)$ (see text
for details). The open points indicating total baryonic density at
$z=0$ \citep{FP:04}, $z=2.5$ \citep{Dun:03}, and $z=3$ \citep{Fuk:98}.
The filled circles are the neutral gas mass density (H+He) in DLA systems.
 \label{fig5}}
\end{figure*}

We turn now to related properties of the DLA population, in order to try to
explain their apparently low average SFRs at high redshift. The mass
densities in stars, $\rho_*$, and DLA neutral gas, $\rho_{\rm DLA}$,
are compared in Figure~\ref{fig5}, along with the total baryonic density for
reference. It can be seen that the gas currently detected in DLA systems
evolves very little over the entire range $0<z\lesssim5$, even including the
factor of $\sim2$ evolution between $z=0$ and $z=0.6$. Also shown in this
Figure is a simple estimate of how the neutral gas mass density,
$\rho_{\rm gas}$, in galaxies may evolve
\citep[see also the closed box model of][]{PF:95}. This is obtained by
assuming that the total $\rho_*(z)+\rho_{\rm gas}(z)$ in galaxies at all
epochs is the same as the local value, $\rho_*(0) + \rho_{\rm gas}(0)$
\citep[see also][their equation~5]{SP:99}.
We thus have $\rho_{\rm gas}(z)=\rho_*(0)+\rho_{\rm gas}(0)-\rho_*(z)$.
Interestingly, the density of neutral
gas in DLA systems under this assumption appears to be a small fraction of
the total $\rho_{\rm gas}$ in galaxies at redshifts above $z\approx1$.
This difference is exacerbated if the stellar density evolution from
\citet{Col:01} is assumed. This picture is quite different from the
widely accepted assumption that DLA systems host the majority of the neutral
gas at all redshifts. Although some recent work suggested that sub-DLA
systems may contribute almost $50\%$ to $\rho_{\rm DLA}$ at $z>3.5$
\citep{Per:03}, this now appears not to be the case \citep{Pro:04}.

How else can we explain this apparent discrepancy? It is unlikely that the
difference can be attributed to DLAs missed due to dust obscuration in
quasars, since the limit on the contribution from missed DLAs is at most
around a factor of two \citep{Ell:01,Ell:04}. One possibility suggested
in \S\ref{rhoz} is that there is significant mass in very high column
density regions, corresponding to the high SFR regions in LBG or
submillimeter galaxy systems, but having such small physical cross-sectional
areas that they are too rare to have been detected in DLA surveys, which have
so far led to the discovery of a few hundred DLAs (and about 100 have
metallicity measurements).
This is consistent with the SFR intensity distribution function illustrated
by \citet{Lan:02}, which suggests that DLA systems only contribute to the low
SFR surface density end. In turn this implies that the high SFR surface
density objects correspond to very high column density neutral plus molecular
gas ($\approx10^{22}$ to $\approx10^{25}$ atoms cm$^{-2}$), which is up
to {\em three orders of magnitude higher than the highest measured DLA
column densities}. These column densities are consistent with the high gas
and SFR surface densities seen in local starburst nuclei \citep{Ken:98}.

We can make a rough estimate of the requirements which would
lead to us missing significant metal-rich neutral and molecular
gas in current DLA surveys. The data suggest that we need to
find $\approx10$ times more gas, and that it needs to have a
metallicity $>0.1$ times solar near $z\approx4$ and greater than
solar metallicity near $z\approx1$. At the same time, since several
hundred DLAs have been identified (but only about 100
have measured metallicities), this rare metal-rich population of
DLA absorbers should have an incidence which is $<1$\% of the known
DLA incidence. In current DLA surveys the mean neutral hydrogen
column density of a DLA is $\approx 10^{21}$ atoms\,cm$^{-2}$
\citep{Rao:05b}. Therefore, the required population of metal-rich DLAs should
have $\langle N_{\rm HI} \rangle > 10^{24}$r$_{0.01}$ atoms\,cm$^{-2}$
where r$_{0.01}=1$ is the current conservative upper limit on the ratio of
the incidence of metal-rich DLAs to DLAs relative to 1\%. We note that
a single metal-rich absorber with an effective radius of $\approx100$\,pc
(e.g., typical of giant neutral hydrogen clouds in galaxies and a factor of
a few larger than typical giant molecular clouds which are the sites for SF)
has a cross section which is $\approx10^4$ times smaller than known DLAs,
which typically have effective radii of $\approx10$\,kpc. Thus, if
r$_{0.01}=1$, only with a survey that identified $\approx10^3$ DLAs would we
detect $\approx10$ absorbers from this putative metal-rich population with
$\langle N_{\rm HI+H_2} \rangle \approx 10^{24}$ atoms\,cm$^{-2}$. The
rare $z\approx0.6$ absorber in Q0218+357 \citep{Mul:05} may be such an example.

Another alternative, of course, is that the closed box assumption of a fixed
$\rho_*(z)+\rho_{\rm gas}(z)$ in galaxies at all redshifts may be too
simplistic. The complexities of galaxy evolution include continuous gas infall
to galaxy potential wells, and subsequent cooling to convert high redshift
ionized gas in the IGM to lower redshift neutral gas in galaxies. This
mechanism may contribute to the relative constancy of
$\rho_{\rm DLA}$, with the neutral gas being replenished as star formation
progresses in these systems. Such a scenario, however, might very well
still lead to inconsistencies with observed stellar metallicities.

In any case, the low value of $\rho_{\rm DLA}$ compared to the stellar mass
density in luminous galaxies is one of the primary reasons for the disparity
in $\rho_Z$ shown in Figure~\ref{fig4}. The combination of low average
metallicity plus low gas mass density naturally leads to a space density of
metals significantly lower than in the luminous galaxy population.

Finally, consider the properties of the observed population of DLAs in
isolation. Within the hierarchical scenario, results on DLA evolution might
be explained by assuming the population is dominated by dwarf galaxies,
especially at high redshift.
Dwarf systems have low SFRs on average \citep[e.g.,][]{Hop:02},
or a late onset of SF \citep{Hop:01,Mat:98}, as well as column densities
within the DLA regime \citep[e.g., the Small Magellanic Cloud,][]{Tum:02}.
They also have low metallicities which are maintained at
a low level despite SF events through galactic winds or other outflow
\citep[e.g.,][]{MF:99}. The DLAs appear to contribute significantly to
the SFH only for $z\lesssim 0.6$, which would correspond to the epoch at
which the late onset SF would occur. At least some
moderately low redshift DLAs ($z\approx 0.5$) are likely to be near
$L^*$ disk galaxies \citep{Rao:03}, and this is also consistent with the
DLA systems appearing to contribute more significantly to the SFH around
this redshift.

It is also possible that DLAs, whether dwarfs or not, could exist in the
outskirts of the same halos as the more centrally located LBGs. This
would naturally account for the column density of DLAs being lower than
the LBGs, and would also result in a higher interception probability for
the DLAs. The low metallicity of the DLA systems could also be naturally
explained as a result of metallicity gradients in the halo \citep{Chen:05}.
This would further account for the DLAs not dominating the neutral gas
content, because most of the gas would be contained in the central
part of the halo with the LBGs. This scenario then makes a strong
prediction that LBGs should be seen in the vicinity of all or
most DLAs. The relative scarcity of such alignments \citep[only one has
been reported in the literature to date,][despite significant efforts to
image DLA galaxies]{Mul:05} suggests that this scenario is unlikely to
be representative of the majority of DLA systems. Perhaps a more
reasonable scenario is that outlined by \citet{Mo:98,Mo:99}, in which
DLA systems are the progenitors of disk and low-surface brightness
galaxies, evolving in predominantly low galaxy density environments
\citep{Mo:98}, while LBGs are the progenitors of massive early type
galaxies, evolving in high galaxy density environments \citep{Mo:99}.

The nature of DLA systems has also been explored through simulations.
Results from \citet{ON:05} and \citet{Nag:04}, for example, indicate that
the average masses for these systems are likely to be low, perhaps around
$10^9\,M_{\odot}$, which is consistent with a galaxy population having
low integrated levels of SFR. The SFR distribution for DLAs was
explored by \citet{ON:05} who find a broad range, spanning $10^{-6}$
to $100\,M_{\odot}$\,yr$^{-1}$, but with a mean value of
$0.01\,M_{\odot}$\,yr$^{-1}$ for DLAs lying within $0<z<1$. They conclude
that DLAs are dwarf systems with typically low SFRs. This is consistent
with the SFH seen in the current study, and also suggests a
plausible reason for the low $\rho_Z$ seen in the DLA population. Given
the outflow mechanism for maintaining the low metallicity in dwarfs, the
stellar populations in the DLA systems (as well as their ISM) would be
expected to show the low metallicities characteristic of nearby dwarf systems.
This is indeed seen in a local ($z\approx 0.01$) dwarf galaxy identified
as a DLA \citep{RSL:04, RSL:05}. These results are not inconsistent with
the high SFRs estimated from [OIII] emission in three high-z DLAs recently
detected by \citet{Wea:05}. These systems are likely to be at the bright end
of the distribution of DLA properties, consistent with the high SFR tail seen
in the simulations of \citet{ON:05}. This is further suggested by a search
for H$\alpha$ emission from DLAs at $z>2$, in which \citet{Bun:99} find
upper limits on the SFRs of a few tens of solar masses per year.

\section{Summary}
\label{summ}
Using the local relationship between gas and SFR surface densities
from \citet{Ken:98} we derive the SFH of the DLA population.
At low redshift ($z\lesssim 0.6$) the DLAs appear to contribute
significantly to the SFH, but they have a much lower contribution
at high redshift. This indicates that at high redshift the majority of the
DLA systems are unlikely to be the same population as the LBGs, or at
least they sample a very different luminosity regime.
An exploration of the evolution of stellar and metal mass densities,
as well as a comparison of $\rho_{\rm DLA}$ with the inferred total
gas mass density, suggests that the DLA population may be dominated by
dwarf like systems with low average SFRs or a late onset of SF.
It also suggests that the DLAs identified so far may not account for the
majority of the neutral gas at high redshift.

\acknowledgements
We thank the referee for comments that have improved this study. We
thank Art Wolfe for providing a copy of his latest results prior to
publication, and for discussion of his interpretation of those results.
We also thank Andy Bunker, Brigitte K{\"o}nig, Regina Schulte-Ladbeck,
and Ken Nagamine for valuable discussion. AMH acknowledges support
provided by grants NSF CAREER AST9984924 and NASA LTSA NAG58546.
SMR and DAT acknowledge support from STScI grant HST-GO-09382.01-A.

\begin{deluxetable}{cccc}
\tablewidth{0pt}
\tablecaption{Vertices of the envelope in Figure~\protect\ref{fig1}.
 \label{tab1}}
\tablehead{
\colhead{Vertex} & \colhead{$\log(1+z)$} & \colhead{$z$} & \colhead{$\log(\dot{\rho}_*)$}
}
\startdata
A & 0.00 & 0.00 & -2.02 \\
B & 0.48 & 2.02 & -0.90 \\
C & 0.68 & 3.79 & -0.90 \\
D & 0.83 & 5.76 & -1.05 \\
E & 0.83 & 5.76 & -0.35 \\
F & 0.36 & 1.29 & -0.35 \\
G & 0.00 & 0.00 & -1.52 \\
\enddata
\tablecomments{Vertices are labelled ABCDEFG moving anticlockwise about the
envelope, starting from the lower $z=0$ vertex.}
\end{deluxetable}

\begin{thebibliography}{}
\bibitem[Brinchmann \& Ellis(2000)]{Bri:00}
   Brinchmann, J., Ellis, R. S. 2000, \apj, 536, L77
\bibitem[Bruzual \& Charlot(2003)]{BC:03}
   Bruzual, G., Charlot, S. 2003, \mnras, 344, 1000
\bibitem[Bunker et al.(1999)]{Bun:99}
   Bunker, A. J., Warren, S. J., Clements, D. L., Williger, G. M.,
   Hewett, P. C. 1999, \mnras, 309, 875
\bibitem[Calura \& Matteucci(2004)]{CM:04}
   Calura, F., Matteucci, F. 2004, \mnras, 350, 351
\bibitem[Chen et al.(2005)]{Chen:05}
   Chen, H.-W., Kennicutt, R. C., Rauch, M. 2005, \apj, 620, 703
\bibitem[Cole et al.(2001)]{Col:01}
   Cole, S., et al. 2001, \mnras, 326, 255
\bibitem[Conti et al.(2003)]{Con:03}
   Conti, A., et al. 2003, \aj, 126, 2330
\bibitem[Curran et al.(2004a)]{Cur:04a}
   Curran, S. J., Webb, J. K., Murphy, M. T., Carswell, R. F.
   2004a, \mnras, 351, L24
\bibitem[Curran et al.(2004b)]{Cur:04b}
   Curran, S. J., Murphy, M. T., Pihlstr{\"o}m, Y. M., Webb, J. K.,
   Bolatto, A. D., Bower, G. C. 2004b, \mnras, 352, 563
\bibitem[Dickinson et al.(2003)]{Dic:03}
   Dickinson, M., Papovich, C., Ferguson, H. C., Budav{\'a}ri, T.
   2003, \apj, 587, 25
\bibitem[Dunne et al.(2003)]{Dun:03}
   Dunne, L., Eales, S. A., Edmunds, M. G. 2003, \mnras, 341, 589
\bibitem[Ellison et al.(2001)]{Ell:01}
   Ellison, S. L., Yan, L., Hook, I. M., Pettini, M., Wall, J. V.,
   Shaver, P. 2001, A\&A, 379, 393
\bibitem[Ellison et al.(2004)]{Ell:04}
   Ellison, S. L., Churchill, C. W., Rix, S. A., Pettini, M.
   2004, \apj, 615, 118
\bibitem[Fukugita et al.(1998)]{Fuk:98}
   Fukugita, M., Hogan, C. J., Peebles, P. J. E. 1998, \apj, 503, 518
\bibitem[Fukugita \& Peebles(2004)]{FP:04}
   Fukugita, M., Peebles, P. J. E. 2004, \apj, 616, 643
\bibitem[Hopkins(2004)]{Hop:04}
   Hopkins, A. M. 2004, \apj, 615, 209
\bibitem[Hopkins et al.(2001)]{Hop:01}
   Hopkins, A. M., Irwin, M. J., Connolly, A. J. 2001, \apjl, 558, L31
\bibitem[Hopkins et al.(2002)]{Hop:02}
   Hopkins, A. M., Schulte-Ladbeck, R. E., Drozdovsky, I. O.
   2002, \aj, 124, 862
\bibitem[Kennicutt(1998)]{Ken:98}
   Kennicutt, R. C., Jr. 1998, \apj, 498, 541
\bibitem[Lanzetta et al.(2002)]{Lan:02}
   Lanzetta, K. M., Yahata, N., Pascarelle, S., Chen, H.-W.,
   Fern{\'a}ndez-Soto, A. 2002, \apj, 570, 492
\bibitem[Lanzetta et al.(1995)]{Lan:95}
   Lanzetta, K. M., Wolfe, A. M., Turnshek, D. A. 1995, \apj, 440, 435
\bibitem[Mac Low \& Ferrara(1999)]{MF:99}
   Mac Low, M.-M., Ferrara, A. 1999, \apj, 513, 142
\bibitem[Madau et al.(1996)]{Mad:96}
   Madau, P., Ferguson, H. C., Dickinson, M. E., Giavalisco, M.,
   Steidel, C. C., Fruchter, A. 1996, \mnras, 283, 1388
\bibitem[Mateo(1998)]{Mat:98}
   Mateo, M. 1998, \araa, 36, 435
\bibitem[Miralda-Escud{\'e}(2005)]{ME:05}
   Miralda-Escud{\'e}, J. 2005, \apjl, 620, L91
\bibitem[Mo et al.(1998)]{Mo:98}
   Mo, H. J., Mao, S., White, S. D. M. 1998, \mnras, 295, 319
\bibitem[Mo et al.(1999)]{Mo:99}
   Mo, H. J., Mao, S., White, S. D. M. 1999, \mnras, 304, 175
\bibitem[Muller(2005)]{Mul:05}
   Muller, S. 2005, in ``Probing Galaxies Through Quasar Absorption Lines,"
   IAU Symposium 199, eds P. R. Williams, C. Shu, B. M{\'e}nard, Shanghai
\bibitem[Nagamine et al.(2004a)]{Nag:04a}
   Nagamine, K., Cen, R., Hernquist, L., Ostriker, J. P.,
   Springel, V. 2004a, \apj, 610, 45
\bibitem[Nagamine et al.(2004b)]{Nag:04}
   Nagamine, K., Springel, V., Hernquist, L. 2004b, \mnras, 348, 435
\bibitem[Okoshi \& Nagashima(2005)]{ON:05}
   Okoshi, K., Nagashima, M. 2005, \apj, (in press; astro-ph/0412561)
\bibitem[Pei \& Fall(1995)]{PF:95}
   Pei, Y. C., Fall, S. M. 1995, \apj, 454, 69
\bibitem[P{\'e}roux et al.(2003)]{Per:03}
   P{\'e}roux, C., McMahon, R. G., Storrie-Lombardi, L. J., Irwin, M. J.
   2003, \mnras, 346, 1103
\bibitem[Prochaska \& Herbert-Fort(2004)]{Pro:04}
   Prochaska, J. X., Herbert-Fort, S. 2004, \pasp, 116, 622
\bibitem[Rao \& Turnshek(1998)]{RT:98}
   Rao, S. M., Turnshek, D. A. 1998, \apj, 500, L115
\bibitem[Rao \& Turnshek(2000)]{RT:00}
   Rao, S. M., Turnshek, D. A. 2000, \apjs, 130, 1
\bibitem[Rao et al.(2003)]{Rao:03}
   Rao, S. M., Nestor, D. B., Turnshek, D. A., Lane, W. M., Monier, E. M.
   Bergeron, J. 2003, \apj, 595, 94
\bibitem[Rao et al.(2005a)]{Rao:05a}
   Rao, S. M., Prochaska, J. X., Howk, J. C., Wolfe, A. M.
   2005a, AJ, 129, 9
\bibitem[Rao et al.(2005b)]{Rao:05b}
   Rao, S. M., Turnshek, D. A., Nestor, D. B. 2005b, (in preparation)
\bibitem[Rudnick et al.(2003)]{Rud:03}
   Rudnick, G., et al. 2003, \apj, 599, 847
\bibitem[Ryan-Weber et al.(2003)]{Rya:03}
   Ryan-Weber, E. V., Webster, R. L., Staveley-Smith, L.
   2003, \mnras, 343, 1195
\bibitem[Ryan-Weber et al.(2005)]{Rya:05}
   Ryan-Weber, E. V., Webster, R. L., Staveley-Smith, L.
   2005, \mnras, 356, 1600
\bibitem[Salpeter(1955)]{Sal:55}
   Salpeter, E. E. 1955, \apj, 121, 161
\bibitem[Salucci \& Persic(1999)]{SP:99}
   Salucci, P., Persic, M. 1999, \mnras, 309, 923
\bibitem[Schulte-Ladbeck et al.(2005)]{RSL:05}
   Schulte-Ladbeck, R. E., K{\"o}nig, B., Miller, C. J., Hopkins, A. M.,
   Drozdovsky, I. O., Turnshek, D. A., Hopp, U. 2005, \apjl, (submitted)
\bibitem[Schulte-Ladbeck et al.(2004)]{RSL:04}
   Schulte-Ladbeck, R. E., Rao, S., Drozdovsky, I. O., Turnshek, D. A.,
   Nestor, D. B. 2004, \apj, 600, 613
\bibitem[Tumlinson et al.(2002)]{Tum:02}
   Tumlinson, J., et al. 2002, \apj, 566, 857
\bibitem[Turnshek et al.(2001)]{Tur:01}
   Turnshek, D. A., Rao, S. M., Nestor, D., Lane, W., Monier, E., Bergeron, J.
   Smette, A. 2001, \apj, 553, 288
\bibitem[Weatherley et al.(2005)]{Wea:05}
   Weatherley, S. J., Warren, S. J., M{\o}ller, P., Fall, S. M.,
   Fynbo, J. U., Croom, S. M. 2005, \mnras, (in press; astro-ph/0501422)
\bibitem[Wolfe (2005)]{Wol:05}
   Wolfe, A. M. 2005, in ``Probing Galaxies through Quasar Absorption Lines,"
   IAU Symposium 199, eds P. R. Williams, C. Shu, B. M{\'e}nard, Shanghai
\bibitem[Wolfe et al.(2003b)]{Wol:03b}
   Wolfe, A. M., Gawiser, E., Prochaska, J. X. 2003b, \apj, 593, 235
\bibitem[Wolfe et al.(2003a)]{Wol:03a}
   Wolfe, A. M., Prochaska, J. X., Gawiser, E. 2003a, \apj, 593, 215
\bibitem[York et al.(1986)]{York:86}
   York, D. G., Dopita, M., Green, R., Bechtold, J. 1986, \apj, 311, 610
\end{thebibliography}
\end{document}